\documentclass{XrU2005}
\usepackage{graphicx}

\title{Extending the X-ray luminosity function of AGN to high redshift}

\author[1]{J. Silverman}

\author[2]{P. Green}
\author[2]{W. Barkhouse}
\author[2]{R. Cameron}
\author[2]{M. Kim}
\author[2]{D.-W. Kim}
\author[2]{B. Wilkes}
\author[1]{G. Hasinger}
\affil[1]{Max-Planck-Institut f\"ur extraterrestrische Physik, D-84571 Garching, Germany}
\affil[2]{Harvard-Smithsonian Center for Astrophysics, 60 Garden Street, Cambridge, MA 02138}
\author[]{the full ChaMP project}

\begin{document}


\maketitle

\begin{abstract}

X-ray surveys of the extragalactic universe are now able to detect
significant numbers of AGN out to high redshift ($z\sim5$).  We
highlight some results from the $Chandra$ Multiwavelength Project
(ChaMP) to measure the X-ray luminosity function out to these early
epochs.  At $z>3$, we show that the comoving space density of
luminous ($log~L_X>44.5$) AGN has a behavior similar to the optical
QSO luminosity function.  With a newly compiled sample of AGN from
ChaMP supplemented with those from additional surveys including the
$Chandra$ Deep fields, we present a preliminary measure of the
luminosity function in the hard (2-8 keV) band.  With 37 AGN at $z>3$,
we continue to see a decline in the space density at high redshift
over a wider range in luminosity.  We discuss the need to identify a
larger sample of obscured AGN at high redshift to determine if an
early epoch of hidden supermassive black hole growth occurred.

\end{abstract}

\section{Introduction}

Our present understanding of the evolution of accreting supermassive
black holes (SMBHs) over vast lengths of cosmic time comes from our
measure of the luminosity function (i.e. the number undergoing a luminous 
phase within a specific comoving volume as a function of luminosity
and redshift).  Energy production through mass accretion onto SMBHs
allows us to identify these sites that manifest themselves
observationally as the familiar Active Galactic Nuclei (AGN).  The
luminosity function provides a key constraint to discern the
underlying black hole mass and accretion rate distributions as a
function of redshift from the observed global properties of AGN.  An
accurate assessment of these should elucidate the mechanisms
(i.e. galaxy mergers and/or self-regulated growth) that are
instrumental in their formation and evolution.

To date, an enormous effort has been undertaken to measure the
luminosity function over the wide range in luminosity spanned by AGN
out to high redshift.  The bright end has been well established to
$z\sim5$ by optical imaging surveys which primarily select QSOs using
a multi-color criteria.  The most dramatic feature found is the rise
and fall of the comoving space density with peak activity at
$z\sim2.5$.  With an unprecedented sample of 23,338 in the 2dF QSO
Redshift Survey (2QZ), Croom et al. (2004) convincingly show a
systematic decrease in luminosity (pure luminosity evolution; PLE)
from $z=2$ to the present with very few bright QSOs in the local
universe in compliance with past surveys (e.g. Boyle et al. 1988) .
This fading of the luminous QSO population may be due to a decrease in
the mass accretion rate (e.g.  Cavaliere \& Vittorini 2000) that
appears to be linked to the evolution of the galaxies in which they
reside (e.g. Croton et al. 2005; Di Matteo et al. 2005).  The dropoff
in the space density at $z>3$ (Warren, Hewett \& Osmer 1994; Schmidt,
Schneider \& Gunn 1995; Fan et al. 2001; Wolf et al. 2003) may be
indicative of either the detection of the onset of accretion onto
young SMBHs or a high-redshift population that has been missed,
possibly under a veil of obscuration (Fabian 1999).  Excessive amounts
of dust and gas may be ubiquitous in galaxies at early epochs as a
result of enhanced galaxy formation.

It has been evident for quite some time that optical surveys miss a
significant fraction of the AGN population.  With the luminosity
function being a steeply declining function of luminosity, optical
techniques fail to find the majority of AGN due to spectroscopic
dilution by host galaxy starlight.  Though current techniques do show
considerable improvement (Richards et al. 2005), they still fail to account for
many low luminosity AGN selected by other means.  Of equal
significance, many AGN (e.g. Seyerts 2s) are underrepresented due to
intrinsic dust obscuration and can only be adequately selected in the
low redshift universe (e.g. Hao et al. 2005).  As pertinent for this
proceeding, current models and recent observations continue to
attribute the bulk of the Cosmic X-ray Background (CXRB), the
previously unresolved X-ray emission, to various types of obscured AGN
(see Brandt \& Hasinger 2005 for a review).

In the current era of $Chandra$ and $XMM-Newton$, X-ray surveys are
detecting AGN and QSOs not only enshrouded by heavy obscuration
($N_H>10^{22}$ cm$^{-2}$) but those at high redshift ($z>3$) with
statistics comparable to the optical surveys due to their superb
resolving power between 0.5 to 10 keV.  Previous observatories such as
$EINSTEIN$ and $ROSAT$ were limited to the soft band which biases
samples against absorption.  With 1-2 Msec observations of the
$Chandra$ Deep Field North (CDF-N; Alexander et al. 2003), Deep Field
South (CDF-S; Rosati et al. 2002) and Lockman Hole (Hasinger et al.\
2001), $\sim89\%$ of the hard (2-8 keV) CXRB has been resolved into
point sources (Moretti et al.\ 2003). Many of the hardest
serendipitous sources found so far arise in optically unremarkable
bright galaxies (e.g. Barger et al.\ 2003b; Tozzi et al.\ 2001;
Mainieri et al. 2002), which can contain very heavily obscured AGN.

A more robust luminosity dependent evolutionary scheme (see Miyaji,
Hasinger \& Schmidt 2000) has emerged in recent measures of the X-ray
luminosity function (XLF).  With the inclusion of absorbed AGN from
$Chandra$, $XMM-Newton$ and $ASCA$ surveys, lower luminosity AGN are
clearly more prevalent at lower redshifts ($z<1$) than those of high
luminosity that peak at $z\sim2.5$.  This behavior has been well
substantiated using hard (2-8 keV) X-ray selected surveys of all types
of AGN (Cowie et al. 2003; Barger et al. 2003a; Ueda et al. 2003;
Fiore et al. 2003; Silverman et al. 2005b; La Franca et
al. 2005). Using a highly complete soft (0.5-2.0 keV) band selected
sample of over 1000 type 1 AGN, Hasinger, Miyaji \& Schmidt (2005) show
that this LDDE model accurately fits the data and shows a gradual
shift of the peak in the comoving space density to lower redshifts
with declining luminosity.  In contrast to an evolution model in which
SMBHs at low redshifts have sub-Eddington accretion rates, this
behavior may be evidence for the growth of lower mass black holes
emerging in an 'anti-hierarchical' or ``cosmic downsizing'' fashion
while accreting near their Eddington limit (e.g. Merloni et al. 2004;
Marconi et al. 2004).

Even though this model is quite convincing, there are remaining
uncertainties in our current measure of the XLF.  (1) A significant
number of X-ray sources in the recent surveys with $Chandra$ and
$XMM-Newton$ are not identified.  (2) Barger et al. (2005) demonstrate
that the XLF can be fit equally well by a PLE model at $z<1.2$.  These
models only begin to substantially differ at low luminosities
(i.e. below the break) and higher redshifts where statistics are quite
low with most being provided by the CDF-N and CDF-S observations.  New
moderate depth surveys such as the Extended Chandra Deep Field South
(E-CDF-S; Lehmer et al. 2005) and the Groth strip (Nandra et al. 2005)
will provide additional AGN at these luminosities and redshifts but
await optical followup.  (3) Behavior of the AGN population at
redshifts above the peak is still not accurately constrained.  We have
presented preliminary evidence (Silverman et al. 2005b) for a similar
evolution of luminous X-ray selected QSOs to those found in the
optical surveys with a decline in the comoving space density at $z>3$
but these AGN are mainly unobscured (type 1).  In consideration of
similar findings from radio selected surveys (Wall et al. 2005) that
are sensitive to obscured QSOs, our current understanding of the
evolution of high redshift AGN may endure.

In this proceeding, we present some results from the ChaMP that extend
our knowledge of the X-ray luminosity function at high redshift.  We
also present some preliminary results from our effort to measure the
X-ray luminosity function in the hard (intrinsic 2-8 keV) band.  These
new results will be further elaborated and expanded to include model
fits using a maximum likelihood method in a near future ChaMP paper
(Silverman et al. in preparation).  To date, the limited numbers of
X-ray selected AGN at $z>3$ have constrained current measures (Ueda et
al. 2003; Barger et al. 2005) to lower redshifts.  Due to the rarity
of luminous high redshift AGN, such an endeavor requires a survey that
covers a wide enough area to sufficient depths.  The $Chandra$
Multiwavelength Project (ChaMP) is carrying out such a survey.  To
improve the dynamic range ($L_X-z$ coverage) of our sample, we include
those AGN found in the CLASXS (Yang et al. 2004; Steffen et al. 2004),
the deep surveys with Chandra (i.e. CDFN, CDFS) and XMM-Newton Lockman
Hole (Hasinger et al. 2001) that have published catalogs with a fair
sample of low luminosity ($42<log~L_X<44$) AGN out to $z\sim5$.

Throughout this work, we assume H$_{\circ}=70$ km s$^{-1}$ Mpc$^{-1}$,
$\Omega_{\Lambda}=0.7$, and $\Omega_{\rm{M}}=0.3$.

\section{Chandra Multiwavelength Project (ChaMP)}

The ChaMP (Kim et al. 2004a,b; Green et al. 2004) is carrying out an
extragalactic X-ray survey encompassing 10 deg$^{2}$ using
serendipitous detections in archival {\em Chandra} fields.  We have
chosen 27 of the 135 ChaMP fields (2.0 deg$^{2}$) for which we have
acquired extensive followup optical imaging and spectroscopy.  The
deepest observations have exposure times that are sensitive to sources
with $f_{0.5-2.0\rm{keV}}>5\times10^{-16}$ erg cm$^{-2}$ s$^{-1}$.  A
full description of the ChaMP image reduction and analysis pipeline
XPIPE can be found in Kim et al. 2004a.  With our 4m MOSAIC optical
imaging, we are able to identify counterparts to the {\em Chandra}
sources down to $r^{\prime}\sim25$ (Green et al. 2004).  We acquire
optical imaging in three ($g^{\prime}, r^{\prime},$ and $i^{\prime}$)
Sloan Digital Sky Survey (SDSS) filters.  Optical colors provide
preliminary source classification and crude photometric redshifts.
Optical spectroscopic followup currently focuses on identifying
counterparts with $r^{\prime}<23.0$ for which spectra can be acquired
on a 4-6m (i.e. MMT, Magellan, WIYN, CTIO Blanco) class telescope.  To
date, we have spectroscopically classified a sample of $\sim450$ AGN.

\section{Co-moving space density of luminous AGN}

Using AGN from the ChaMP and those from additional surveys (CDF-N,
CDF-S, $ROSAT$), we have measured the co-moving space density type 1,
highly luminous (log~$L_{\rm X}>44.5$) AGN.  In Figure 1, the space
density rises from the present epoch to a peak at $z\sim2.5$ and then
declines at $z>3$ (Silverman et al. 2005b).  This behavior is similar
to that of optically selected samples (Fan et al. 2001; Schmidt,
Schneider \& Gunn 1995; Wolf et al. 2003).  These results are further
substantated by the recent space densities of soft X-ray selected AGN
reported by Hasinger, Miyaji \& Schmidt (2005) and radio selected QSOs
(Wall et al. 2005).  These results all support the scenario that SMBHs
are rapidly growing at the early epochs possibly related to enhanced
galaxy formation.

\begin{figure}
\centering
\includegraphics[width=1.0\linewidth]{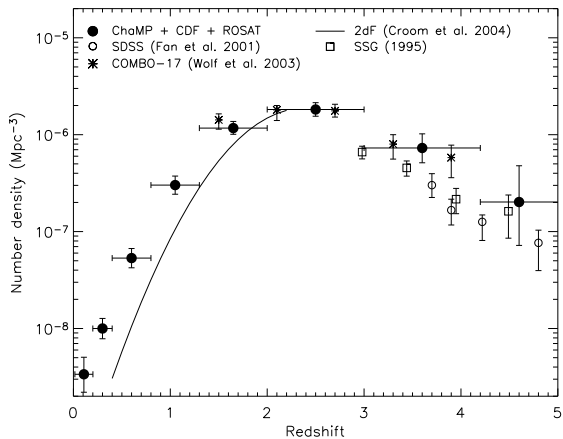}
\caption{Co-moving space density of 217 {\em Chandra} + {\em ROSAT} AGN
selected in the soft (0.5--2.0 keV) band with log~$L_{\rm X}>44.5$
compared to the optical surveys.  The optical space densities have
been scaled to match the X-ray points at $z=2.5$ for ease of
comparison.}
\end{figure}

\section{Hard X-ray luminosity function}

We are also generating a X-ray luminosity function in the 2-8 keV band
to minimize the effect of intrinsic absorption.  Similar to Barger et
al. (2005), we use the observed soft X-ray band for AGN selection
above a specific redshift, in our case this is $z=3$ where we measure
the rest frame energies above 2 keV.  The hard band (2-8 keV)
detections are used to construct a low redshift ($z<3$) sample of AGN.
This enables our selection to be less affected by absorbing columns up
to $N_H\sim10^{23}$ cm$^{-2}$.  The use of the soft (0.5-2.0 keV) band
allows us to take advantage of Chandra's high collecting area at low
energies to detect faint, high redshift AGN.  We compile both a sample
of AGN from 896 hard and 1232 soft selected X-ray sources found in 27
ChaMP fields.  These fields were chosen to have a limiting flux
($f_{0.5-2.0~keV}\sim10^{-15}$ erg s$^{-1}$ cm$^{-2}$) capable of
detecting high redshift AGN, quality optical imaging in the
$r^{\prime}$ band for the low redshift ($z<3$) sample, $i^{\prime}$
for the higher redshift sample, and a substantial amount of optical
spectroscopic followup.  We further restrict our hard band catalog to
those with $r^{\prime}<22.5$ and the soft band sources to
$i^{\prime}<23.5$ since our spectroscopic followup campaign has
provided a significant number of spectroscopic redshifts brighter than
these magnitude limits.  To construct a pure AGN sample, we require
the rest frame 2.0--8.0 keV luminosity (uncorrected for intrinsic
absorption) to exceed 10$^{42}$ erg s$^{-1}$.  These selection
criteria yield a sample of 360 AGN (Figure 2) from the ChaMP with 18
at $z>3$ that represents half of the known X-ray selected AGN at high
redshift.

\subsection{Covering the luminosity-redshift plane}

\begin{figure}
\centering
\includegraphics[width=1.0\linewidth]{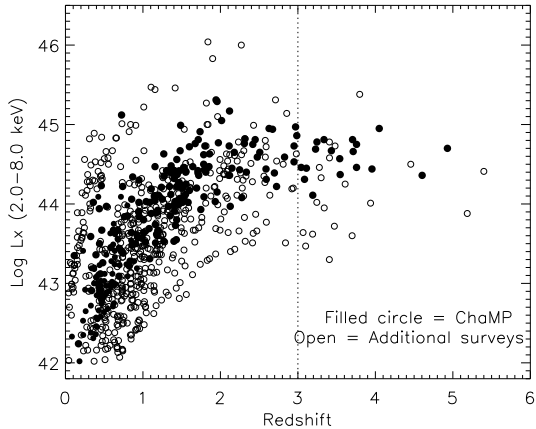}

\caption{Luminosity-redshift distribution of AGN from the ChaMP
(filled circles) and those from the CDF-N, CDF-S, XMM/Lockman
Hole. AMSSn and CLASXS (open circles).}
\label{lx_z}
\end{figure}

We supplement the ChaMP AGN with those from the following hard X-ray
surveys to measure the XLF over the wide range of luminosity and
redshift spanned by AGN: $Chandra$ Deep Field North (CDF-N; Alexander
et al. 2003, Barger et al. 2003b), $Chandra$ Deep Field South (CDF-S;
Giacconi et al. 2002; Szokoly et al. 2004; Zheng et al. 2004),
$Chandra$ Large Area Synoptic X-ray Survey (CLASXS; Yang et al. 2004,
Steffen et al. 2004), $XMM-Newton$ Lockman Hole (Brunner et al. in
preparation), ASCA Medium Sensitivity Survey (AMSSn; Akiyama et
al. 2003).  Both a hard and soft X-ray catalog are constructed as
implemented for the ChaMP sample.  Multi-band optical imaging in these
fields allows us to convert optical magnitudes to the SDSS filter
system ($r^{\prime}$,$i^{\prime}$) used by the ChaMP.  The
distribution in luminosity and redshift of the full AGN sample is
shown in Figure~\ref{lx_z}.  As evident, the deep surveys (i.e. CDF-N
and CDF-S) are instrumental to detect low luminosity ($log~L_X<44$)
AGN above a redshift of 1.5.  In contrast, the ChaMP supplies the more
luminous $log~L_X>44$ AGN underrepresented in the deep fields.

\subsection{Method}

We measure the differential X-ray luminosity function
(d$\Phi/\rm dlogL$) expressed in Equation~\ref{eq:phi_def} where $N$
is the number of AGN per unit comoving volume ($V$) per $log~L_{\rm
x}$ as a function of luminosity $L_{\rm x}$ and redshift $z$.

\begin{equation}
\frac{d\Phi(L_{\rm x},z)}{{\rm dlog}\,L_{\rm x}}=\frac{{\rm d}^{2}\it{N}}{{\rm d}V\,{\rm d}logL_{\rm x}}(L_{\rm x},z)
\label{eq:phi_def}
\end{equation}

\noindent This function is assumed to be continuous over the range in
luminosity and redshift for which our survey is sensitive.  The
differential luminosity (d$logL$) is expressed as a logarithm (base
10) due to the 4 orders of magnitude spanned by our sample.

Here, we estimate the XLF in fixed luminosity and redshift bins using
the 1/V$_a$ method (Schmidt 1968; Avni \& Bahcall 1980).  For each
$L-z$ bin, the value of the XLF is a sum (Equation~\ref{eq:va}) of the
contribution from each AGN falling within this specific bin.

\begin{equation}
\frac{d\Phi(L_{\rm x},z)}{{\rm dlog}~L_{\rm x}}=\frac{1}{\Delta {\rm log} L_X}\sum_{i=1}^{N} C_i \frac{1}{V_{a,i}}
\label{eq:va}
\end{equation}

We must apply a correction factor ($C_i$) for each AGN to account for
the incompleteness in our optical spectroscopic identifications.  This
factor is reciprocal of the fraction of identified sources ($f_{\rm
ID}$) at X-ray fluxes and optical magnitudes comparable to each
source.  The accessible volume $V_a$ is a function of both X-ray and
optical limiting fluxes.  Further details on our measure of both $V_a$
and $f_{\rm ID}$ are given in Silverman et al. (2005b).  We estimate 1$\sigma$
errors based on a Poisson distribution due to the small number of
objects per redshift bin.

Since we have a sample selected in two different energy bands, the
survey area over which we are sensitive to an individual AGN depends
on the observed flux.  The sky coverage for AGN at $z<3$ depends on
their hard flux while higher redshift AGN depend on their soft flux.

\subsection{Preliminary results}

\begin{figure*}
\centering
\hspace{4cm}
\includegraphics[angle=90,width=0.5\linewidth]{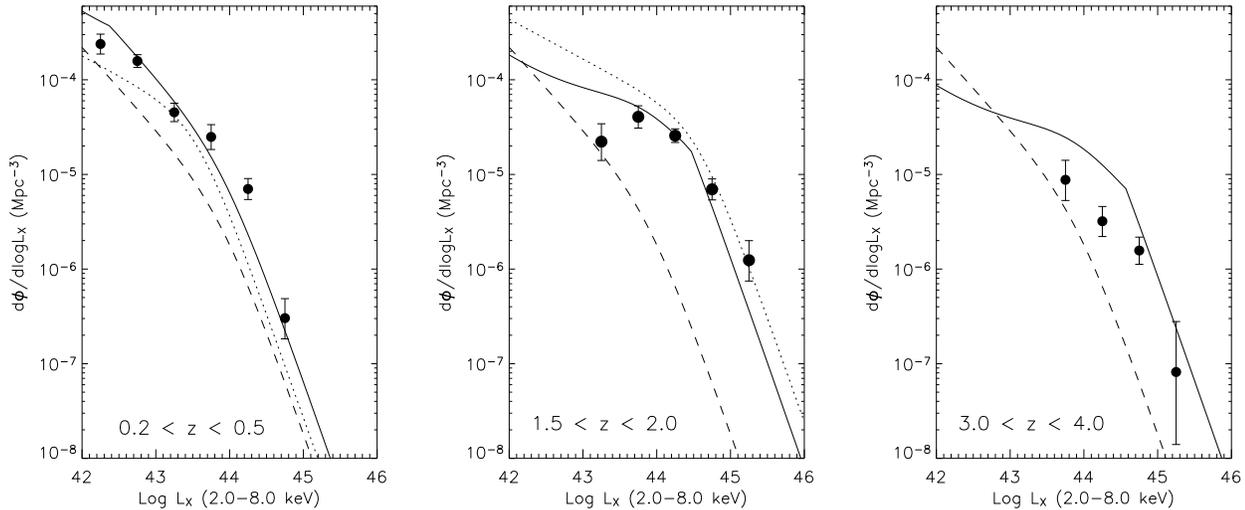}

\caption{Binned XLF ($1/V_a$ method) in three redshift intervals.  The
LDDE model with best fit parameters from Ueda et al. (2003) are
plotted at the median redshift (solid line) and $z=0$ (dashed line).
We extrapolate the model to $z>3$ to illustrate the significance of
the decrease in normalization of the XLF at high redshift.  The pure
luminosity evolution model of Barger et al. (2005) is shown by the
dotted line and extrapolated up to $z=2$.}
\label{xlf_bin}
\end{figure*}

In Figure~\ref{xlf_bin}, we plot the binned XLF in three redshift
intervals to highlight the strong evolutionary trends.  The luminosity
function over the full redshift interval $0<z<5.5$ will be presented
as mentioned in an upcoming paper (Silverman et al., in preparation).
The XLF at $z<2.0$ is well approximated by the familiar double
powerlaw where we have sufficient statistics to cover the full
luminosity range.  There is a strong increase in characteristic
luminosity for the entire sample to $z=2$.  To evaluate our XLF, we
have plotted the best fit pure luminosity evolution (PLE) model from
Barger et al. (2005) and luminosity-dependent density evolution (LDDE)
model of Ueda et al. (2003).  Due to the similarity of both these
models at $z<1.5$, it is difficult to determine which one better
matches our data.  There may be some evidence of a flattening of the
slope at the faint end ($log~L_X<44$) for $1.5<z<2.0$ that may agree
with a LDDE model but any solid evidence of this requires a more
sophisticated analysis.

We find that the Ueda model provides a good fit to our data for $z<3$.
At higher redshifts ($3<z<4$), we see a decline in the overall numbers
of AGN at all luminosities, similar to the decline present in the soft
selected samples (Figure 1).  With this new sample, we are able to
further constrain the slope and normalization of the XLF at these high
redshifts.

\begin{figure}
\centering
\includegraphics[width=1.0\linewidth]{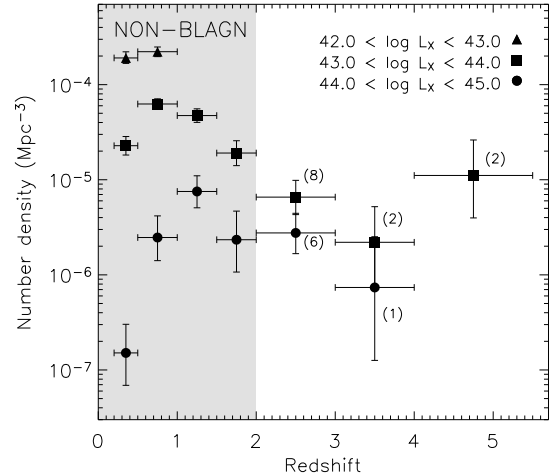}
\caption{Comoving space density of AGN without broad optical emission lines
in their optical spectra. The white area highlights the redshift range
where samples are quite limited with the number given in parenthesis.}
\label{nonblagn}
\end{figure}

With knowledge of the optical spectral properties of these AGN, we can
investigate the evolution of those with a hidden broad emission line
region.  With the exception of those at low luminosity, these AGN are
usually obscured and many can be optically classified as type 2 in
compliance with unification models.  In Figure~\ref{nonblagn}, we plot
their comoving space density in three luminosity classes.  First, it
is quite evident from the error bars in Figure 4 that the sample is
much smaller.  Even with this limited sample, it appears qualitatively
that more luminous AGN peak at earlier epochs ($z\sim2$) than those of
lower luminosity ($z<1$) similar to the behavior of the type 1 AGN
population (Hasinger, Miyaji \& Schmidt 2005).  This may imply that
type 1 and type 2 AGN evolve in a similar manner and an obscured epoch
of AGN formation is not evident.  Larger samples are clearly needed to
substantiate any such claims.

\section{Conclusion}

Even with a further understanding of the evolution of X-ray selected
AGN at high redshift, we are continuing to identify a larger sample of
ChaMP sources through deep optical spectroscopy.  We must reduce the
number of unidentified sources at faint optical magnitudes
($r^{\prime}>22.5$).  We expect that this optically faint population
will include a significant number of the luminous ($log~L_X>44$) type
2 QSOs that have not been detected in large numbers in the ChaMP.  With
$\sim20$ $z>4$ candidates based on optical colors, we may be able to
significantly improve the statistics at these redshifts for which ChaMP
has already found four to date.

\end{document}